\documentclass[fleqn,10pt,x11names]{wlscirep}

\usepackage[utf8]{inputenc}
\usepackage{multicol}
\usepackage{amssymb}
\usepackage{hyperref}
\usepackage{amsmath}
\usepackage{graphicx}
\usepackage[normalem]{ulem}
\usepackage{makecell}
\usepackage{csquotes}

\usepackage[backend=biber, style=alphabetic, sorting=nyt]{biblatex}
\usepackage{titlesec}

\title{Navigating the challenges in creating complex data systems: a development philosophy}

\author[1,6,+,*]{S{\"{o}}ren Dittmer}
\author[1,+,*]{Michael Roberts}
\author[1]{Julian Gilbey}
\author[1]{Ander Biguri}
\author[2]{AIX-COVNET Collaboration}
\author[3]{Jacobus Preller}
\author[4]{James H.F. Rudd}
\author[5]{John A.D. Aston}
\author[1]{Carola-Bibiane Sch{\"{o}}nlieb}

\affil[1]{Department of Applied Mathematics and Theoretical Physics, University of Cambridge, Cambridge, UK}
\affil[2]{A list of authors and their affiliations appears at the end of the paper}
\affil[3]{Addenbrooke’s Hospital, Cambridge University Hospitals NHS Trust, Cambridge, UK.}
\affil[4]{Department of Medicine, University of Cambridge, Cambridge, UK}
\affil[5]{Department of Pure Mathematics and Mathematical Statistics, University of Cambridge, Cambridge, UK}
\affil[6]{ZeTeM, University of Bremen, Bremen, Germany}

\affil[+]{these authors contributed equally to this work}
\affil[*]{corresponding authors sd870@cam.ac.uk and mr808@cam.ac.uk}

\begin{abstract}
In this perspective, we argue that despite the democratization of powerful tools for data science and machine learning over the last decade, developing the code for a trustworthy and effective data science system (DSS) is getting harder. 
Perverse incentives and a lack of widespread software engineering (SE) skills are among many root causes we identify that naturally give rise to the current systemic crisis in reproducibility of DSSs.
We analyze why SE and building large complex systems is, in general, hard. Based on these insights, we identify how SE addresses those difficulties and how we can apply and generalize SE methods to construct DSSs that are fit for purpose.
We advocate two key development philosophies, namely that one should incrementally grow -- not biphasically plan and build -- DSSs, and 
one should always employ two types of feedback loops during development:
one which tests the code's correctness and another that evaluates the code's efficacy.
\end{abstract}

\begin{document}

\maketitle

\begin{multicols}{2}

Machine learning is in a reproducibility crisis~\cite{haibe2020transparency,pineau2021improving,baker20161}. We argue that a primary driver is poor code quality, having two root causes: poor incentives to produce good code and a widespread lack of Software Engineering (SE) skills.
The crisis also demonstrates that Data Science Systems (DSS) can, and will, fail silently if no continual verification infrastructure exists throughout their development~\cite{recipe,jama.2019.11954, acs.orglett.9b03216}.

We consider two questions important to all data scientists. Firstly, why is it so hard to build complex systems? Here we blame the intrinsic fragility and sheer number of components in modern DSSs, holding for the code and data involved. Therefore, we reason that the development of DSSs must follow Gall's law -- one can not build complex systems; one has to grow them~\cite{gall1975general}. Note that this aligns well with agile development in modern SE. Secondly, we ask, how can we write trustworthy and effective DSSs? We argue that having two types of feedback loops in place is a critical necessity: one to assess the DSS's correctness and another to assess its efficacy.

In fact, we believe the crisis to be a natural corollary of an environment in which data scientists develop complex DSSs without growing them nor establishing a careful and continual assessment of their code's correctness. While incorrect code can be computationally reproducible  -- i.e., rerunning the code produces identical results -- for replicability and general reproducibility using independent implementations, correctness is crucial. Maybe even more crucial, reusability -- standing on the shoulders of giants -- demands correctness; indeed, without it every downstream task inherits the lack of correctness and reproducibility.

\section*{The problem: a Cambrian explosion}
Until relatively recently, statisticians had a monopoly on data analysis. They were, and are, highly trained to appreciate the intricate relationships and biases in data and to use relatively \emph{simple} methods (in the best sense of the word) to analyze the data and fit models to it. Data collection was often done under their guidance to ensure biases were understood, documented, and mitigated.

Nowadays, data is ubiquitous and often claimed to be the new oil.
However, real-world datasets often resemble more of an oil spill, containing a plethora of unknown (and often unknowable) biases. Without sufficient statistics and SE skills, the development of a DSS tends to lead to the following implications:
\begin{align*}
    \text{Big Data} &\Rightarrow \text{Messy Data} \Rightarrow \text{Big Code}\\ &\Rightarrow \text{Messy Code}  \Rightarrow \text{Incorrect Conclusions}
\end{align*}
The radical increase in the scale and availability of data has led to an equally radical paradigm shift in its use. Data scientists build complex systems on top of complex, biased, and generally incomprehensible data. To do this, they are the consumers of many more software tools than classical statisticians. As a user of many tools, it is naturally more vital to know how to interface with them and less possible to understand their internal workings. Hence the underlying software must be trustworthy; one has to assume it is almost bug-free, with any remaining bugs being insignificant to any conclusions.

Expressing and structuring an analysis plan in code is the bedrock for all data science projects and due to these many tools, modern data scientists must write increasing amounts of custom `glue code' when developing DSSs. However, SE is a challenging discipline, and building on vast unfamiliar  
codebases often leads to unexpected consequences. 

\textit{Both} from the data and algorithm's perspective, this paradigm shift resembles a Cambrian explosion in the quantity and intrinsic complexity of data and code.

\section*{Why is the problem challenging?}
In this section, we want to discuss some significant challenges data scientists face when developing a correct and effective DSS. Some of these challenges are due to human nature, whereas others are of a technical nature.

\subsection*{Challenge 1: Missing SE skills}
Most data scientists only learn to write small codebases, whereas SE focuses on working with large codebases. As mentioned above, code is the interface to many data science tools, and SE is the discipline of organizing interfaces methodically. For this paper, \textbf{we define SE as the discipline of managing the complexity of code and data with interfaces as one of its primary tools}~\cite{parnas1972criteria}. While many SE practices focus on enterprise software and do not trivially apply to all components of DSSs, it is our conviction that SE methodologies must play a more prominent role in future data science projects.

\subsection*{Challenge 2: Correctness and efficacy}
\begin{figure*}
\centering
    \begin{tabular}{c||c|c} 
      & Not correct & Correct \\ [0.5ex] 
     \hline\hline
     \hline
     Not efficacious & \makecell{You do not know whether\\ your idea is bad.\\ Try to achieve correctness,\\ it might give you\\ efficacy too.} & You need a new idea. \\ 
     \hline
     Efficacious & \makecell{You do not know whether\\ your idea works.\\ Try to make the system\\ correct or analyze why\\ your system is effective.} & \makecell{\includegraphics{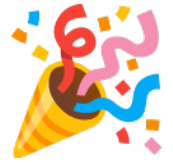}}
    \end{tabular}

    \caption{You need both: correctness \& efficacy.}
    \label{fig:table}
\end{figure*}

A DSS must work correctly, i.e., it does what you think it does. It also must be efficacious, i.e., produce relevant and usable predictions. Without SE, following earlier arguments, this tends to lead to the following implications:
\begin{align*}
    &\text{Multiple Experiments} \Rightarrow \\
    &\ \text{Messy Code} \Rightarrow \text{Incorrect Conclusions}
\end{align*}
So why do we truly need correctness \textit{and} efficacy for a trustworthy high-performing model? Firstly, as mentioned, a published, executable code can provide computational reproducibility, but repeatability requires correctness. Secondly, while an incorrect DSS can be efficacious due to a lucky bug, it is uninterpretable and hard to modify. \textbf{Without correctness, it is impossible to understand, interpret, or trust the outputs of and conclusion based on a DSS.} See Figure~\ref{fig:table} for visualization of why we need correctness and efficacy.

\subsection*{Challenge 3: Perverse incentives in academia}
Software engineers, industrial data scientists, and academic data scientists produce different products within wildly different incentive structures. Software engineers are rewarded for creating high-performing, well-documented, and reusable codebases; data scientists are rewarded based on their DSS outputs. Like software engineers, industrial data scientists are rewarded based on the system's usefulness to the company. Academic data scientists, however, aim to use their results to write marketable papers to further their field, apply for grants, and enhance their reputation.

For academia, there is a conflict between short-term and long-term incentives.

Academic careers are peripatetic in nature and most positions are temporary for early career researchers, who tend to be those developing DSSs. Therefore, in the short-term it is rewarding to publish papers quickly, and give less attention to reusability of the codebase, as careful reusable development leads to delayed gratification. The short-term academic incentive structure might even discourage producing and publishing code comprehensible for a broad audience to avoid getting `scooped' by competitors.

In the long-term however, a clear incentive to develop reusable DSSs is that this increases the probability that the paper will become influential and be well cited. For example, if two similar papers are published, but only one provides good code, it is almost certain that future papers will compare directly to this one. Over time, this will dramatically (and multiplicative) separate the popularity of the two papers. Not having this incentivised however leads to an enormous value destruction for society. 

However, the grant system gives one potential mechanism to resolve the perverse encourage realisation of these long-term incentives by encouraging proposals that involve the development of DSSs must involve resources for the construction of reusable and deployable codebases. 

Interestingly, this is not a new phenomenon; Knuth~\cite{knuth1984literate} discussed it in the 1980s when he was advised to publish \TeX's source code.
However, if a field's incentive structure and goals are misaligned, see, e.g., the positive publication bias~\cite{barnett2019examination,van2021significance}, the path of least resistance easily wins the upper hand.


\subsection*{Challenge 4: Short-circuits}
The democratization of powerful data analysis and machine learning tools allows for short-circuits as keen amateurs can develop complex DSSs relatively quickly. This is not to say that using powerful publicly available tools or short-circuits are inherently bad. On the contrary, if every practitioner was writing private versions of common tool kits, this would be a major source of bugs. 

However, powerful tools reduce the accidental complexity, not the intrinsic complexity of DSS. Thus, they make it easier to build complex systems having a high intrinsic complexity. This intrinsic complexity is extremely hard to manage, especially because it often has hidden subtleties.


\subsection*{Challenge 5: Teams vs. individual work}
Working in a team, e.g., on a codebase, can be extremely powerful, but without the proper training or organizational structure, it can also produce massive inefficiencies and errors -- teams being complex systems themselves. Software engineers are often highly trained in agile teamwork methodologies, e.g., SCRUM~\cite{sutherland2014scrum, fowler2001agile}. They also know how to harness the benefits of infrastructure, such as version control, continuous integration pipelines, and pair programming. Academic data scientists tend only to possess informal training in these teamwork-enabling tools.

\subsection*{Challenge 6: Bridging the academia-industry gap}
Data science projects in industry and academia have many similarities. However, besides the already discussed incentive differences, there are also key distinctions in the DSS development environment. Due to a larger SE culture, industry embraces the idea that high-quality code is obligatory for maintainable DSSs; academia is often simply not interested in, or rewarded for, maintainability. In academia, the incentives promote a strong throwaway mentality towards code. Many DSSs never break out from research groups and, usually, there is no incentive for long-term maintenance. Finally, academia has virtually no feedback loop for code quality. High-quality code is neither a pre-requisite for most publications nor utilized for assessing career performance.

\subsection*{Challenge 7: Training a DSS is costly}
A change in a DSS can require costly and lengthy retraining, either to check how it changes the outcome or to check that it does not change the outcome. For this reason, seemingly minor fixes and improvements, as well as code cleanups, might not happen at all.

\subsection*{Challenge 8: Long-term maintenance}
Even a small DSS is often sufficiently complex that the number of dependencies on other packages or codes can number in the dozens or hundreds. As complex systems are inherently fragile, a minor change in one of the dependencies can lead to a (potentially silent) failure of the entire DSS. This is one of many reasons long-term code maintenance is costly or simply not possible. While there are many countermeasures to facilitate computation reproducibility, e.g., publishing Python/Anaconda environments and test suites, they do not ensure future reusability within larger DSS.

\subsection*{Summary of the challenges}
Many researchers have a systemic lack of awareness that SE is integral to modern data science. This not only translates to a lack of formal training but also into a perverse incentive structure. Considering human nature, it is somewhat surprising that there is good academic data science code at all. These perverse incentives cause a colossal loss of opportunity to create value!

DSSs must be \emph{both} correct and efficacious. The potential unleashed by the useability of modern data science tools has enabled significant progress but also the development of many seemingly efficacious but incorrect systems.

Industry is inherently better at developing high-quality code as their code must integrate with infrastructure, teams, and deployment platforms. Academia lacks such guard rails; code development is often myopic.

\section*{Using software engineering to grow complex systems}
Every programmer can write small codebases, but larger ones require SE to perform both correctly and be maintainable~\cite{farley2021}. So, why is it generally hard to build complex systems from scratch?

Complex systems tend to consist of many highly interconnected components which are fragile to small perturbations. In the case of a codebase, these perturbations could be simple typos that, with luck, produce a syntax error. Otherwise, a simple typo can subtly alter the outcome in unknown ways and lead to dramatic and unexpected consequences.

Gall's law~\cite{gall1975general} states that complex systems cannot be built; they can only be grown, i.e., we should not plan the entire DSS in advance, implement it and then evaluate the code. Instead, one should use small incremental steps following a not-too-detailed plan, never deviating far from a working system. Gall's law should be of great value to data scientists as we argue that growing an $n$-component system can reduce the maximal build complexity from $O(n^2)$ to $O(n)$. 

Although one can often decompose complex systems into predominantly simple components, the sheer number of them quickly turns the interacting simple components into a complex whole. If one wants to build a system with $n$ components, there are up to $O(n^2)$ interactions between them, giving $O(n^2)$ potential failure points (assuming that each component works correctly). SE has developed two leading solutions to this ``$O(n^{2})$-problem": software architecture and agile development~\cite{bass2003software, farley2021}.

\begin{figure*}
    \centering
    \includegraphics[trim=25 25 25 25,clip,width=\textwidth]{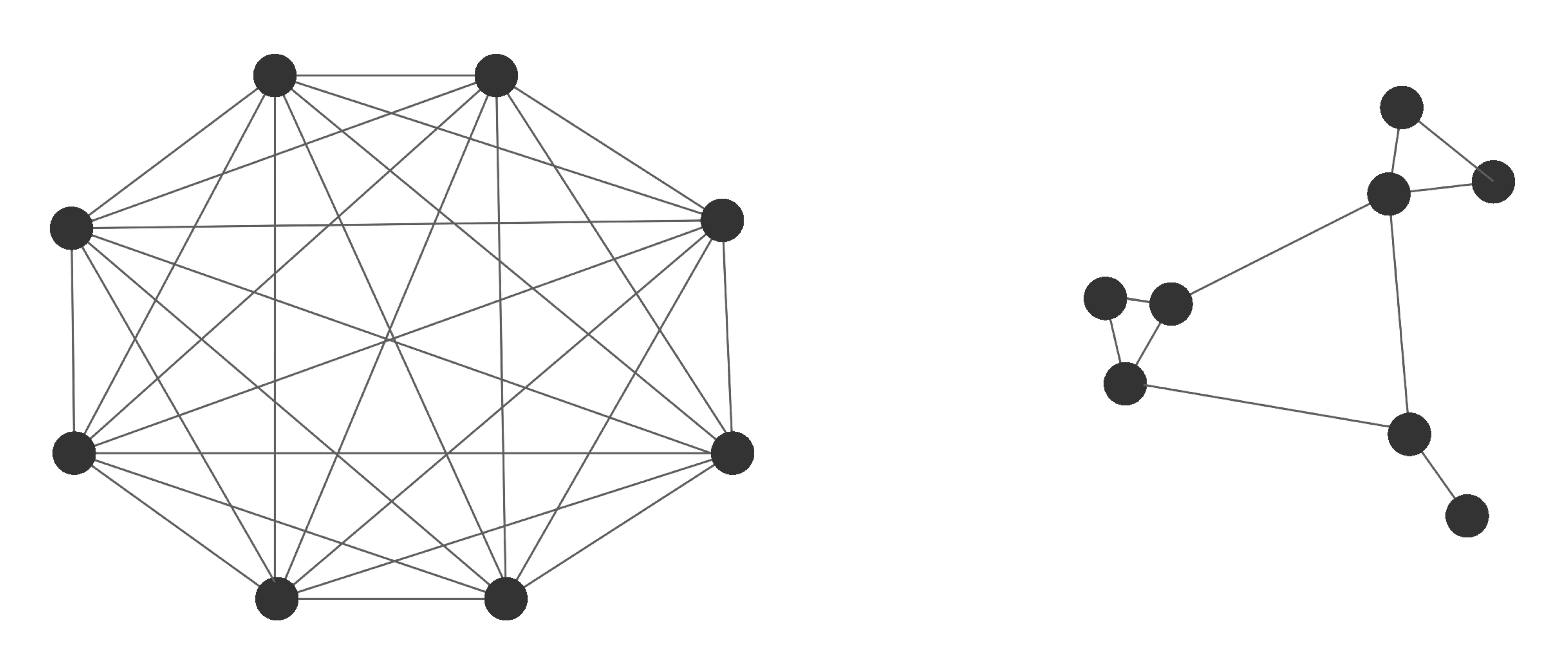}
    \caption{The graph on the left is a fully connected graph illustrating a bad software architecture. The graph on the right is sparsely, and mostly locally, connected, demonstrating a better architecture~\cite{watts1998collective,valverde2003hierarchical}.}
    \label{fig:graph_network}
\end{figure*}

\textbf{Software Architecture.} Well-established code development principles are a critical component of SE. One key principle is the separation of concerns, which splits the software into different components, each handling a single isolated concern and possessing a simple, complexity-hiding interface~\cite{parnas1972criteria}. These components are, in turn, formed by connecting lower-level isolated components. Designing the software architecture in this manner reduces the graph spanned by the different components from a potentially densely connected graph with $O(n^{2})$ connections to a sparse graph with fewer connections, i.e., far fewer potential failure points. It is also advantageous to have a sense of locality in the code and graph such that components are preferably locally connected, see Figure~\ref{fig:graph_network} for a visualization.

\begin{figure*}
    \centering
    \includegraphics[width=\textwidth]{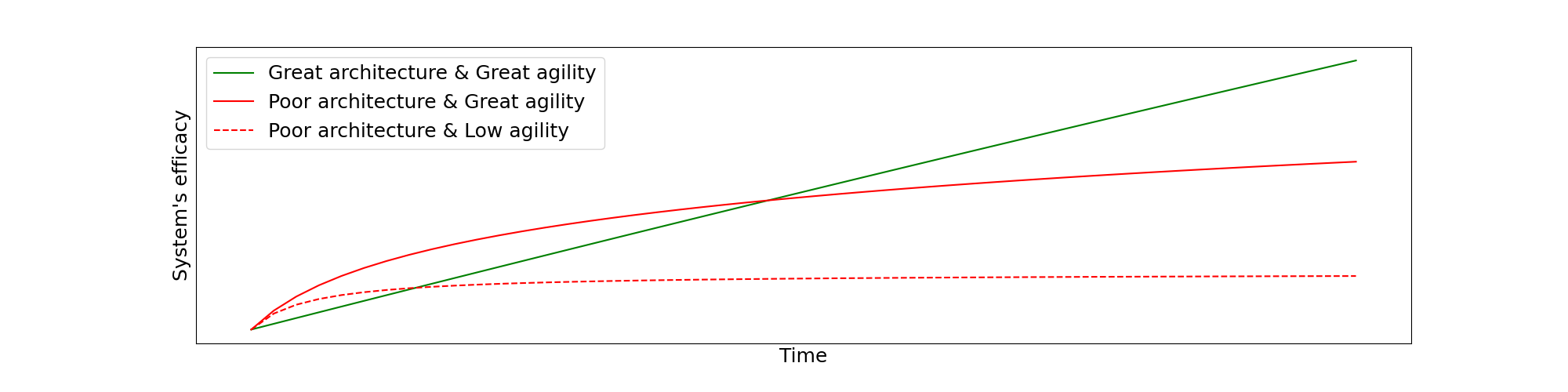}
    \caption{Some SE practices require delayed gratification, as one has to sacrifice in the short-term for long-term progress.}
    \label{fig:delayed_gratification}
\end{figure*}

\textbf{Agile Development.} Modern SE tends to follow an agile approach, where one grows software incrementally, adding or changing one component at a time, so there is always a working system. One only has to consider how this new component interacts with the existing $n$ components. This reduces the $O(n^{2})$ potential failure points to $O(n)$ possible failure points at each step. In the end, this reduces the ``build complexity" from $O(n^{2})$ to $O(n)$ when the complex system is grown. See Figure~\ref{fig:delayed_gratification} for a visualization.

Another crucial component of SE is the infrastructure used, e.g., continuous integration pipelines and version control systems, along with test suites for codebases. There is empirical evidence that following these principles leads to improvements in SE~\cite{48455, radziwill2020accelerate}. While one might not need a software engineer to develop a DSS, the same principles apply, and one certainly needs software engineering.

\section*{Corrective action is required}
We now want to discuss how we can combine and generalize these principles into concrete advice. We hope to enable data scientists to write correct and effective code in less time.

\subsection*{Do not build complex systems; grow them}
The approach of planning an entire system and then building it does not work for complex systems and \textbf{we must grow DSSs} to keep the complexity on the order of at most $O(n)$ at each incremental stage, ideally -- enabled by good software architecture -- of $O(1)$. 

Planning is still required to ensure we continually grow our systems towards the desired goal, but it should be highly iterative and alternate with incremental implementation steps. \textbf{Planning not only orients the iterations but also helps to avoid local optima} during the evolution of the DSS. It is valuable to recognize that the future evolution of a complex system is increasingly fuzzy. Planning should follow a multiscale approach with a discount factor on future details. An excellent example, is the comparison of SpaceX's rocket development process against the classical approach \cite{reddy2018spacex,vance2015elon,smith1979shuttle}. The rocket's design was grown, by testing it over many iteratively adapted instantiations, each being a little bit less of a failure than the last one. \cite{perkel2022fix}~argues that this iterative process, embracing the inevitability of errors, must become a deeply appreciated fact during a DSS's development -- and complex (software) system in general.

\subsection*{Testing for data science systems}
Developing a suite of tests alongside a DSS provides a comprehensive correctness feedback loop. \textbf{A good test suite mitigates the fragility} that DSSs -- as complex systems -- inevitably have, e.g., a typo can destroy everything without ever being noticed. However, software engineers often base tests on known example input-output pairs, but knowing these for non-trivial codes may be impossible, e.g., complex numerical code.

\textbf{Property-based testing} can help. While we often do not know a-priori the correct output of a function for a given input, in many cases, mathematics can tell us some properties that the function, or its output, should have. With property-based testing, we use that knowledge, e.g., by creating random inputs to the function and checking whether the property holds for all of them. Python libraries, such as \texttt{pytest}~\cite{pytest} and \texttt{hypothesis}~\cite{Hypothesis} can be utilised for general and property-based testing respectively.

One critical question is: which tests do I have to write to be confident in the correctness of my system? We propose that focusing on the functionality the system will need to provide when deployed is key.
This allows for recursive -- similar to dynamic programming -- thinking about what components of the system one must test and to which degree, to have confidence in the system's functionality when deployed externally.

Within the code, we recommend performing as many integrity checks on the data as possible by implementing tests that check whether the inputs of your system fulfill the assumptions you make about them before it goes into the model~\cite{recipe}. This is hard as we are often unaware of assumptions we make and often forget which assumptions we made, so it makes sense to hard-code them with tests whenever we notice them. One can also do this with dedicated Python libraries like \texttt{pandera}~\cite{pandera}. 
If we expect a variable to be in a particular format, we should write a check that generates an error or warning in case the format is wrong. This approach minimizes the uncertainties in the code by converting assumptions into certainties.

\subsection*{The nature and necessity of feedback loops}
\begin{figure*}
    \centering
    \includegraphics[width=.5\textwidth]{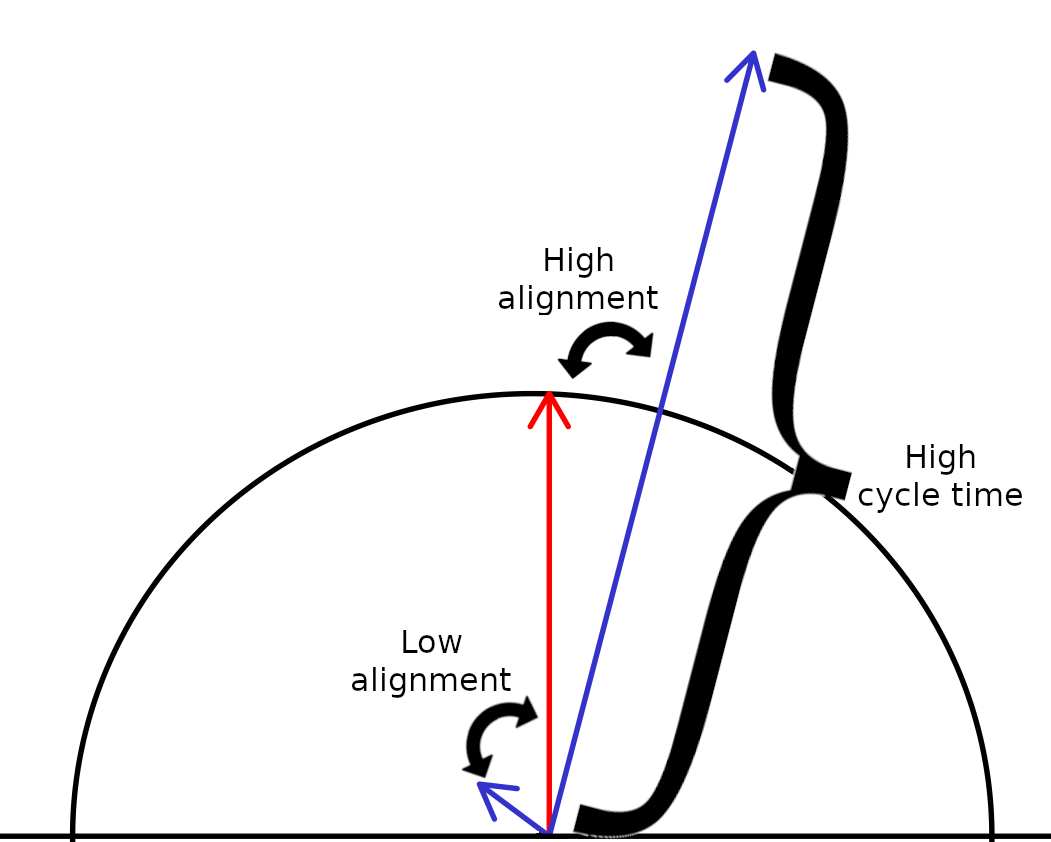}

    \caption{All feedback signals (blue) have an alignment with the goal we have in mind (red) and a cycle time. Often one has to pay for high alignment with high cycle time.}
    \label{fig:feedback}
\end{figure*}

Earlier, we discussed how growing a DSS can reduce the build complexity from $O(n^2)$ to $O(n)$. Incremental, iterative development relies on feedback loops. When establishing a feedback loop, relying on a test function, it is helpful to consider two properties of it: alignment and cycle time, which we also visualize in Figure~\ref{fig:feedback}.

\textbf{Alignment.}  How many assumptions about a given code component are measured by the test function? Aligning the objectives of the test function with our expectations of the code is crucial. If the alignment is not good, the feedback loop cannot give us confidence in the trustworthiness of the component.

\textbf{Cycle time.} How much time (or cost) is required to get the feedback? If we could build a 100\% aligned feedback loop, it is not very helpful if it takes an unreasonable amount of time to run. Ideally, we want a short cycle time to allow for high-frequency feedback.

Writing a test suite is extremely powerful for establishing a feedback loop. Each test run in the test suite returns a measurement on the code, providing a feedback signal about a particular aspect of the code. Responding to the signal of a test suite that has high alignment and low cycle time (running it with high frequency) establishes a strong feedback loop. Readable code is quicker to understand and more reliable, accelerating the cycle time. 

As discussed previously, data scientists should care about \textit{both} the models' efficacy and the code's correctness. Therefore we need feedback loops that measure both. We measure the DSSs efficacy by evaluating our model on a test set and measure its correctness (or trustworthiness) with a test suite and by making the code as readable as possible. 

\subsection*{Software architecture for data science systems}
\begin{figure*}
    \centering
    \includegraphics[width=\textwidth]{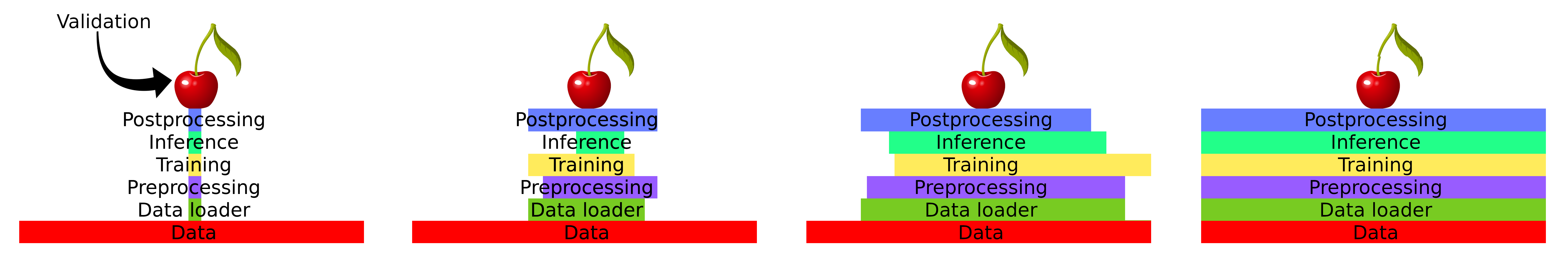}
    \caption{The growth of a DSS over time. You must start by building all layers of your DSS as thin as possible to support the cherry on top as early as possible. This ``steel thread'' is necessary to establish an efficacy feedback loop.}
    \label{fig:cake}
\end{figure*}

Good software architecture helps reduce the number of components connected to an incremental addition to the system, reducing the complexity of building the system. An additional advantage of good architecture is that it dramatically improves the readability of code and keeps the codebase flexible for future developments~\cite{lakshmanan2020machine}.
We argue that the crucial architecture concept for DSSs is the idea of horizontal layers, as shown in Figure~\ref{fig:cake}.

\textbf{Horizontal software layers} are the different components of an analysis pipeline, e.g., data loading, preprocessing, model training, and evaluation. Each of these layers can be seen as a component in the software and should have as few external connections as possible. Each layer is a pocket of complexity, hiding its complexity from the other components in the system.

Feature and model engineering are two of the most important tasks in machine learning, and we can interpret both as asking questions about the data. The ultimate question we often ask is: how well can I predict some labels with given features, particular preprocessing, and a specific model? Answering this requires coding the whole pipeline and then testing how well it works. This is the opposite of the agile approach and forces us to wait long for an efficacy feedback signal. 

Therefore, we recommend organizing the pipeline in horizontal layers but building each layer in a minimalistic incomplete fashion so that the basic connections between the layers are established early on in the project (Figure~\ref{fig:cake}). \textit{We think this point is crucial to quickly establish a tight, highly aligned feedback loop.} Without this feedback loop, even feature preprocessing becomes a potential `fishing expedition' as you can not be sure if it improves the outcome.

This also justifies why using a simple method like linear regression first is a good idea when fitting a predictive model. It is often claimed that you should do this to avoid overfitting, but this is clearly not the case. The key reason is that linear regression is easy to implement and fast to run, enabling you to rapidly establish the first feedback loop with a short cycle time.
In SE terms, this could be called a minimum viable product. For data pipelines, this is often called a steel thread~\cite{playbook} as it bootstraps a stable path that one can gradually extend to build a more complete pipeline.

\subsection*{Data $*$ Code = Complexity$^2$}
 
Data scientists must not only be tolerant of the complexities of code but also of the complexity of data and additional complexities introduced by applying code to data. However well designed the code, the old adage of garbage-in-garbage-out still holds. SE allows for mastery of the complexities of code but combining code with data results in complexity on top of complexity, and constructing a DSS can resemble balancing a stick on top of a stick.

Implementing tests not only for code but also for data~\cite{baumgartner2021, recipe} is clearly a crucial and powerful tool for a data scientist; we want to emphasize again that \textbf{tests turn assumptions into lasting certainties}. It is also crucial to plot the data as often as possible. Looking at plots is a feedback loop with high throughput and can give you high alignment. However, looking at plots is time consuming, i.e., this has a high cycle time. We recommend extracting the relevant information from what you see in the plots and writing tests based on that.

It is critical to appreciate that DSSs often fail silently~\cite{recipe}, as we do not know what the data knows. They should not just churn data but also highlight issues and violated assumptions to the developer. It can seem like everything is working, but in reality, the code is buggy, with machine learning code in general and deep learning code, in particular, failing silently in many unexpected ways. For example, one likely would not notice if early layers of a convolutional neural network's architecture were buggy and their weights not updated during training. 

Questions are asked of the data by running experiments. Like in a good conversation, you must listen to the answers carefully and adapt your future responses and questions accordingly. That does not mean your question generator algorithm has to be greedy, but it has to be iterative. On the one hand, iterative work unlocks the power of feedback loops, which we need when working with complex/real-world data. On the other hand, this requires agility in how you interact with the data, i.e., in your software.

\section*{Conclusion}
We now want to highlight the most valuable and important points.

One has to \textbf{grow DSS} incrementally. Feedback loops are a prerequisite for feature engineering, model development, and everything else. \textbf{Feedback loops allow one to move faster, further, and more confidently.}

\textbf{Correctness and efficacy are different things that require different feedback loops.} The most critical feedback loops for correctness are writing and running a \textbf{test suite} and writing as \textbf{comprehensible code} as possible. The most important point for building a feedback loop for efficacy is to establish it early by growing the \textbf{entire data pipeline as early and thin as possible}.

We note that (almost) no feedback loop is perfectly aligned; still, they are essential. We want to warn that a subtle problem can arise when iterating on misaligned feedback loops: overfitting, also known as Goodhart's law\cite{goodhart1984problems, hoskin1996awful}, stating that every measure which becomes a target ceases to be a good measure. Overfitting is predominantly a problem for efficacy feedback loops. As discussed in~\cite{muller2019tyranny}, people and processes optimizing perverse incentives and misaligned feedback tend to (un-)consciously ``play the system.'' This overfitting, i.e., on the validation set, can happen to the entire DSS, not just the model. While researchers might be aware of this problem when training models, they are often unaware of this problem for DSSs in general. The same countermeasures to model overfitting apply to DSS, e.g., the use of holdout test sets not utilised during the development.

Also, we want to reemphasize that this is a \textbf{socio-technical problem}: despite endeavors like Zenodo or SoftwareX, academia often lacks, or even reverses, incentive structures to create and publish a high-quality DSS. We must, therefore, improve the incentives structure alignment in academia.

\subsection*{Acknowledgements}

There is no direct funding for this study, but the authors are grateful for the EU/EFPIA Innovative Medicines Initiative project DRAGON (101005122) (S.D., M.R., AIX-COVNET, C.-B.S.), the Trinity Challenge BloodCounts! project (M.R., J.G., C.-B.S.), the EPSRC Cambridge Mathematics of Information in Healthcare Hub EP/T017961/1 (M.R., J.H.F.R., J.A.D.A, C.-B.S.), the Cantab Capital Institute for the Mathematics of Information (C.-B.S.). The European Research Council under the European Union’s Horizon 2020 research and innovation programme grant agreement no. 777826 (C.-B.S.), the Alan Turing Institute (C.-B.S.), Wellcome Trust (J.H.F.R.), Cancer Research UK Cambridge Centre (C9685/A25177) (C.-B.S.), British Heart Foundation (J.H.F.R.), the NIHR Cambridge Biomedical Research Centre (J.H.F.R.), HEFCE (J.H.F.R.).
In addition, C.-B.S. acknowledges support from the Leverhulme Trust project on ‘Breaking the non-convexity barrier’, the Philip Leverhulme Prize, the EPSRC grants EP/S026045/1 and EP/T003553/1 and the Wellcome Innovator Award RG98755.
Finally, the AIX-COVNET collaboration is also grateful to Intel for financial support. We also want to thank Jan-Christoph Lohmann, Shaun Griffith, and Jeremy Tang for the helpful comments and discussions.

\subsection*{AIX-COVNET}

Michael Roberts$^{1}$, S{\"{o}}ren Dittmer$^{1,6}$, Ian Selby$^{7}$, Anna Breger$^{1,8}$, Matthew Thorpe$^{9}$, Julian Gilbey$^{1}$, Jonathan R. Weir-McCall$^{7,10}$, Effrossyni Gkrania-Klotsas$^{3}$, Anna Korhonen$^{11}$, Emily Jefferson$^{12}$, Georg Langs$^{13}$, Guang Yang$^{14}$, Helmut Prosch$^{13}$, Jacobus Preller$^{3}$, Jan Stanczuk$^{1}$, Jing Tang$^{15}$, Judith Babar$^{3}$, Lorena Escudero Sánchez$^{7}$, Philip Teare$^{16}$, Mishal Patel$^{16,17}$, Marcel Wassin$^{18}$, Markus Holzer$^{18}$, Nicholas Walton$^{19}$, Pietro Li{\'{o}}$^{20}$, Tolou Shadbahr$^{15}$, James H. F. Rudd$^{4}$, John A.D. Aston$^{5}$, Evis Sala$^{7}$ and Carola-Bibiane Schönlieb$^{1}$.\\

\noindent
${}^{7}$ Department of Radiology, University of Cambridge, Cambridge, UK
${}^{8}$ Faculty of Mathematics, University of Vienna, Austria.
${}^{9}$ Department of Mathematics, University of Manchester, Manchester, UK.
${}^{10}$ Royal Papworth Hospital, Cambridge, Royal Papworth Hospital NHS Foundation Trust, Cambridge, UK
${}^{11}$ Language Technology Laboratory, University of Cambridge, Cambridge, UK.
${}^{12}$ Population Health and Genomics, School of Medicine, University of Dundee, Dundee, UK.
${}^{13}$ Department of Biomedical Imaging and Image-guided Therapy, Computational Imaging Research Lab Medical University of Vienna, Vienna, Austria.
${}^{14}$ National Heart and Lung Institute, Imperial College London, London, UK.
${}^{15}$ Research Program in Systems Oncology, Faculty of Medicine, University of Helsinki, Helsinki, Finland.
${}^{16}$ Data Science \& Artificial Intelligence, AstraZeneca, Cambridge, UK.
${}^{17}$ Clinical Pharmacology \& Safety Sciences, AstraZeneca, Cambridge, UK.
${}^{18}$ contextflow GmbH, Vienna, Austria. 
${}^{19}$ Institute of Astronomy, University of Cambridge, Cambridge, UK. 
${}^{20}$ Department of Computer Science and Technology, University of Cambridge, Cambridge, UK.

\end{multicols}

\medskip
\printbibliography

@book{lakshmanan2020machine,
  title={Machine learning design patterns},
  author={Lakshmanan, Valliappa and Robinson, Sara and Munn, Michael},
  year={2020},
  publisher={O'Reilly Media}
}

@book{farley2021,
  title={Modern Software Engineering: Doing What Works to Build Better Software Faster},
  author={David Farley},
  year={2021},
  publisher={Addison-Wesley Professional}
}

@article{haibe2020transparency,
  title={Transparency and reproducibility in artificial intelligence},
  author={Haibe-Kains, Benjamin and Adam, George Alexandru and Hosny, Ahmed and Khodakarami, Farnoosh and Waldron, Levi and Wang, Bo and McIntosh, Chris and Goldenberg, Anna and Kundaje, Anshul and Greene, Casey S and others},
  journal={Nature},
  volume={586},
  number={7829},
  pages={E14--E16},
  year={2020},
  publisher={Nature Publishing Group}
}

@article{pineau2021improving,
  title={Improving reproducibility in machine learning research: a report from the NeurIPS 2019 reproducibility program},
  author={Pineau, Joelle and Vincent-Lamarre, Philippe and Sinha, Koustuv and Larivi{\`e}re, Vincent and Beygelzimer, Alina and d’Alch{\'e}-Buc, Florence and Fox, Emily and Larochelle, Hugo},
  journal={Journal of Machine Learning Research},
  volume={22},
  year={2021},
  publisher={Microtome Publishing}
}

@article{baker20161,
  title={1,500 scientists lift the lid on reproducibility},
  author={Baker, Monya},
  journal={Nature},
  volume={533},
  number={7604},
  year={2016}
}

@book{gall1975general,
  title={General Systemantics},
  author={Gall, John},
  year={1975},
  publisher={General Systemantics Press}
}

@article{baumgartner2021,
  title={ Ways I Use Testing as a Data Scientist },
  author={Baumgartner, Peter},
  URl={https://www.peterbaumgartner.com/blog/testing-for-data-science/},
  year={2021}
}

@book{sutherland2014scrum,
  title={Scrum: the art of doing twice the work in half the time},
  author={Sutherland, Jeff and Sutherland, Jeffrey Victor},
  year={2014},
  publisher={Currency}
}

@book{bass2003software,
  title={Software architecture in practice},
  author={Bass, Len and Clements, Paul and Kazman, Rick},
  year={2003},
  publisher={Addison-Wesley Professional}
}

@article{fowler2001agile,
  title={The agile manifesto},
  author={Fowler, Martin and Highsmith, Jim and others},
  journal={Software development},
  volume={9},
  number={8},
  pages={28--35},
  year={2001},
  publisher={[San Francisco, CA: Miller Freeman, Inc., 1993-}
}

@techreport{48455,
title	= {2019 Accelerate State of DevOps Report},
author	= {Nicole Forsgren and Dustin Smith and Jez Humble and Jessie Frazelle},
year	= {2019},
URL	= {http://cloud.google.com/devops/state-of-devops/}
}

@misc{radziwill2020accelerate,
  title={Accelerate: Building and Scaling High Performance Technology Organizations.},
  author={Radziwill, Nicole},
  year={2020},
  publisher={Taylor \& Francis}
}

@article{reddy2018spacex,
  title={The spacex effect},
  author={Reddy, Vidya Sagar},
  journal={New Space},
  volume={6},
  number={2},
  pages={125--134},
  year={2018},
  publisher={Mary Ann Liebert, Inc. 140 Huguenot Street, 3rd Floor New Rochelle, NY 10801 USA}
}

@book{vance2015elon,
  title={Elon Musk},
  author={Vance, Ashlee and Sanders, Fred},
  year={2015},
  publisher={HarperCollins}
}

@article{smith1979shuttle,
  title={Shuttle Problems Compromise Space Program: With the shuttle earth-bound, political troubles and cost overruns take off},
  author={Smith, R Jeffrey},
  journal={Science},
  volume={206},
  number={4421},
  pages={910--914},
  year={1979},
  publisher={American Association for the Advancement of Science}
}

@techreport{playbook,
title	= {Data Pipeline Playbook},
author	= {Paul Brabban and Simon Case and Scott Cutts and Claudio Diniz and Lewis Crawford},
year	= {2021},
URL	= {https://data-pipeline.playbook.ee/}
}

@techreport{recipe,
title	= {A Recipe for Training Neural Networks},
author	= {Andrej Karpathy},
year	= {2019},
URL	= {https://karpathy.github.io/2019/04/25/recipe/}
}

@article{parnas1972criteria,
  title={On the criteria to be used in decomposing systems into modules},
  author={Parnas, David L},
  booktitle={Pioneers and their contributions to software engineering},
  pages={479--498},
  year={1972},
  publisher={Springer}
}

@article{jama.2019.11954,
    author = {Aboumatar, Hanan and Wise, Robert A.},
    title = "{Notice of Retraction. Aboumatar et al. Effect of a Program Combining Transitional Care and Long-term Self-management Support on Outcomes of Hospitalized Patients With Chronic Obstructive Pulmonary Disease: A Randomized Clinical Trial. JAMA. 2018;320(22):2335-2343.}",
    journal = {JAMA},
    volume = {322},
    number = {14},
    pages = {1417-1418},
    year = {2019},
    month = {10},
    issn = {0098-7484},
    doi = {10.1001/jama.2019.11954},
    url = {https://doi.org/10.1001/jama.2019.11954},
    eprint = {https://jamanetwork.com/journals/jama/articlepdf/2752474/jama\_aboumatar\_2019\_rx\_190001.pdf},
}

@article{acs.orglett.9b03216,
author = {Bhandari Neupane, Jayanti and Neupane, Ram P. and Luo, Yuheng and Yoshida, Wesley Y. and Sun, Rui and Williams, Philip G.},
title = {Characterization of Leptazolines A–D, Polar Oxazolines from the Cyanobacterium Leptolyngbya sp., Reveals a Glitch with the “Willoughby–Hoye” Scripts for Calculating NMR Chemical Shifts},
journal = {Organic Letters},
volume = {21},
number = {20},
pages = {8449-8453},
year = {2019},
doi = {10.1021/acs.orglett.9b03216},
note ={PMID: 31591889},
URL = {https://doi.org/10.1021/acs.orglett.9b03216},
eprint = {https://doi.org/10.1021/acs.orglett.9b03216}
}

@misc{perkel2022fix,
  title={How to fix your scientific coding errors},
  author={Perkel, Jeffrey M},
  year={2022},
  publisher={Nature Publishing Group}
}

@incollection{muller2019tyranny,
  title={The tyranny of metrics},
  author={Muller, Jerry Z},
  booktitle={The Tyranny of Metrics},
  year={2019},
  publisher={Princeton University Press}
}

@incollection{goodhart1984problems,
  title={Problems of monetary management: the UK experience},
  author={Goodhart, Charles AE},
  booktitle={Monetary theory and practice},
  pages={91--121},
  year={1984},
  publisher={Springer}
}

@article{hoskin1996awful,
  title={The ‘awful idea of accountability’: inscribing people into the measurement of objects},
  author={Hoskin, Keith},
  journal={Accountability: Power, ethos and the technologies of managing},
  volume={265},
  year={1996},
  publisher={International Thomson Business Press London}
}

@article{knuth1984literate,
  title={Literate programming},
  author={Knuth, Donald Ervin},
  journal={The Computer Journal},
  volume={27},
  number={2},
  pages={97--111},
  year={1984},
  publisher={Oxford University Press}
}

@misc{Hypothesis,
  title =            {{H}ypothesis x.y},
  author =     {David R. MacIver},
  year =             {2016},
  howpublished = {\href{https://github.com/HypothesisWorks/hypothesis-python}{\texttt{https://github.com/HypothesisWorks/hypothesis-python}}},
}

@misc{pytest,
  title =        {pytest x.y},
  author = {Krekel, Holger and Oliveira, Bruno and Pfannschmidt, Ronny and Bruynooghe, Floris and Laugher, Brianna and Bruhin, Florian},
  year =         {2004},
  url = {https://github.com/pytest-dev/pytest},
}

@InProceedings{pandera,
  author    = { {N}iels {B}antilan },
  title     = { pandera: {S}tatistical {D}ata {V}alidation of {P}andas {D}ataframes },
  booktitle = { {P}roceedings of the 19th {P}ython in {S}cience {C}onference },
  pages     = { 116 - 124 },
  year      = { 2020 },
  editor    = { {M}eghann {A}garwal and {C}hris {C}alloway and {D}illon {N}iederhut and {D}avid {S}hupe },
  doi       = { 10.25080/Majora-342d178e-010 }
}

@article{barnett2019examination,
  title={Examination of CIs in health and medical journals from 1976 to 2019: an observational study},
  author={Barnett, Adrian Gerard and Wren, Jonathan D},
  journal={BMJ open},
  volume={9},
  number={11},
  pages={e032506},
  year={2019},
  publisher={British Medical Journal Publishing Group}
}

@article{van2021significance,
  title={The significance filter, the winner's curse and the need to shrink},
  author={van Zwet, Erik W and Cator, Eric A},
  journal={Statistica Neerlandica},
  volume={75},
  number={4},
  pages={437--452},
  year={2021},
  publisher={Wiley Online Library}
}

@article{watts1998collective,
  title={Collective dynamics of ‘small-world’networks},
  author={Watts, Duncan J and Strogatz, Steven H},
  journal={nature},
  volume={393},
  number={6684},
  pages={440--442},
  year={1998},
  publisher={Nature Publishing Group}
}

@article{valverde2003hierarchical,
  title={Hierarchical small worlds in software architecture},
  author={Valverde, Sergi and Sol{\'e}, Ricard V},
  journal={arXiv preprint cond-mat/0307278},
  year={2003}
}
\end{document}